\documentclass{IEEEtran}
\usepackage{amssymb}
\usepackage{amsfonts}
\usepackage{amsmath}
\usepackage{algorithm}
\usepackage{algorithmic}
\usepackage{tikz}
\usepackage{graphicx,subfigure,cite,multicol,multirow,diagbox,booktabs,array}
\usepackage{color}
\usepackage{stfloats}
\usepackage{epstopdf}
\UseRawInputEncoding
\DeclareMathOperator*{\argmax}{argmax}
\ifCLASSINFOpdf
\else
\fi
\begin{document}
	\title{ High-performance Power Allocation Strategies for Active IRS-aided Wireless Network\\ }
	
	\author{Yifan~Zhao,~Xuehui~Wang,~Yan~Wang,~Xianpeng~Wang,~Zhilin Chen,~Feng~Shu,~Cunhua~Pan,\\~and~Jiangzhou~Wang~\emph{Fellow},~\emph{IEEE}
		
		\thanks{This work was supported in part by the National Natural Science Foundation of China (Nos.U22A2002, and 62071234), the Hainan Province Science and Technology Special Fund (ZDKJ2021022), the Scientific Research Fund Project of Hainan University under Grant KYQD(ZR)-21008, and the Collaborative Innovation Center of Information Technology, Hainan University (XTCX2022XXC07). \itshape(Corresponding author: Feng Shu).}
		\thanks{Yifan Zhao, Xuehui Wang, Yan Wang, Xianpeng Wang and Zhilin Chen are with the School of Information and Communication Engineering, Hainan University, Haikou 570228, China (e-mail: zyf1001@hainanu.edu.cn;~wangxuehui0503@163.com;~yanwang@hainanu.edu.cn;~zhilin.chen@hainanu.edu.cn.;~wxpeng2016@ hainanu.edu.cn)}
	     \thanks{Feng Shu is with the School of Information and Communication Engineering, and Collaborative Innovation Center of Information Technology, Hainan University, Haikou 570228, China, and also with the School of Electronic and Optical Engineering, Nanjing University of Science and Technology, Nanjing 210094, China (e-mail: shufeng0101@163.com).
		}
		\thanks{Cunhua Pan is with National Mobile Communications Research Laboratory, Southeast University, Nanjing 211111, China. (e-mail: cpan@seu.edu.cn).}
		\thanks{Jiangzhou Wang is with the School of Engineering, University of Kent, CT2 7NT Canterbury, U.K. (e-mail: j.z.wang@kent.ac.uk).}}
	\maketitle
	
	\begin{abstract}
		Due to its intrinsic ability to combat the double-fading effect,  the active intelligent reflective surface (IRS) becomes popular. The main feature of active IRS must  be supplied by power, and  the problem of how to allocate the total power between base station (BS) and IRS to fully explore the rate gain achieved by power allocation (PA) to remove the rate gap between existing PA strategies and optimal exhaustive search (ES) arises naturally. First, the signal-to-noise ratio (SNR) expression is derived to be a function of PA factor $\beta\in [0,~1]$. Then, to improve the rate performance of the conventional gradient ascent (GA), an equal-spacing-multiple-point-initialization  GA (ESMPI-GA) method is proposed. Due to its slow linear convergence from iterative GA, the proposed ESMPI-GA is high-complexity. Eventually,
		to reduce this high complexity, a low-complexity closed-form PA method with  third-order Taylor expansion (TTE) centered at point $\beta_0=0.5$ is proposed.
		Simulation results show that the proposed ESMPI-GA and TTE obviously outperform  existing methods like equal PA.

		\begin{IEEEkeywords}
			Active intelligent reflective surface, power allocation, gradient ascent, equal-spacing-multiple-point-initialization, closed-form, rate performance
		\end{IEEEkeywords}

	\end{abstract}
	
	\section{Introduction}

	With the rapid growth of big data applications such as social media, video streaming, and online games, there is an increased demand for high-speed and stable data transmission. To support the proper operation of these applications, the broader coverage and lower latency should be provided in wireless networks \cite{LYW9424177}.
	In recent years, there are still some difficult problems in the development of wireless communication, such as shadow fading, transmission path loss, energy efficiency, which have slowed down its progress. As a low-cost and low-power
	consumption reflecting device, intelligent reflective surface (IRS) has emerged in response to the needs of wireless communication \cite{LQC10025848,ZN9656117,WW9410435}.

	By intelligently designing the phase shift of each reflecting unit, IRS can build a additionally benefit channel for the incident signal, so that the reflecting signal and the direct signal can be reconfigured  \cite{DLL9998527}. Meanwhile, the performance loss caused by path loss and interference in the direct channel can be significantly improved \cite{LYW9424177}.
	So far, passive IRS has been extensively applied to various wireless communication scenarios.
	Aiming to improve the energy efficiency (EE) of cell-free (CF) network, an IRS-assisted CF network was considered in \cite{LQN9363171}. By utilizing an alternating descent algorithm, the transmit beamforming at the access point and the control matrix at the IRS could be jointly optimized for EE enhancement.
	An IRS-aided multiple-input single-output (MISO) system was proposed in \cite{ZSQ9079457}, while the hardware impairments was taken into account. It was shown that compared to the case of ideal hardware, although the numbers of transmitter antennas and IRS elements went to large scale, the spectral efficiency was finite. Simultaneously, a analytical solution was derived to maximize EE.\cite{YZH10196368} considered an IRS-aided millimeter wave multigroup multicast multiple-input multiple-output (MIMO) communicaion system. To  maximize the sum rate, beamforming vector  and phase shift matrix were jointly optimiazed by combining block diagonalization and manifold method.

	However, there exists a double-fading challenge in the passive IRS-aided cascaded channel, which leads to a limited rate performance enhancement obtained by introducing passive IRS \cite{WQQ10172069}. Given that, active IRS equipped with power amplifiers has appeared to overcome this problem.
	Because of the ability to amplify the power of reflecting signal, the performance loss caused by double fading can be compensated \cite{LYQ10141981}.
	At present, researchers have begun to pay attention to active IRS.
	In \cite{DLL9998527}, the authors demonstrated that the capacity gain of active IRS was substantial by analyzing its asymptotic performance. In addition, an active IRS-aided multi-user MISO (MU-MISO) network was proposed, where transmit beamforming and reflecting coefficient matrix were jointly optimized to maximize sum rate.
	An active IRS-assisted single-input single-output wireless network was considered in \cite{LJ10135161}, where two methods, namely maximum ratio reflecting and selective ratio reflecting, were proposed, which were with closed-form solutions to reflecting coefficient matrix. To further improve the rate performance, an alternately iterative method was presented to solve the problem of maximizing reflected-signal-to-noise ratio.

	When the total transmit power is limited, the rate performance can be significantly improved via power allocation (PA) between transmit signals. PA is a key strategy, which has been well researched in different networks, such as secure spatial modulation (SM) networks \cite{LXY8668810}, secure directional modulation (DM) networks \cite{WSM8315448}, pilot contamination attack (PCA) scheme\cite{WHM9206122}, secure MIMO precoding systems \cite{ST6826572} and passive IRS-assisted decode-forward (DF) relay networks \cite{WXH9887892}.
	 Aiming at maximizing secrecy rate of a secure DM network \cite{WSM8315448}, two PA methods, called general PA and null-space projection PA, were designed to allocate between confidential messages and artificial noise.
	For IRS-assisted DF relay systems in \cite{WXH9887892}, to achieve a goal of maximizing sum rate, successive convex approximation method, maximizing determinant method and a method with rate constraint were respectively proposed to optimize the power factors of two users and DF relay for higher sum rate performance.
	In \cite{cheng2023arXiv231009721C}, the authors focused on the design of PA algorithms for active IRS assisted wireless networks. Given the fixed transmit beamforming vector at BS and IRS phase shift matrix, a Taylor polynomial approximation (TPA) method was proposed, where analytical solutions related to PA factor could be obtained through Taylor approximation. It is particularly noted that the approximate polynomial fit the original function well under the condition of weak direct link from BS to user.

	However, the existing TPA method in \cite{cheng2023arXiv231009721C} is not suitable for some practical scenarios. For example, in a scenario where the direct link between BS and user is strong compared with the reflecting link BS-IRS-User. Moreover, the rate performance gap between the TPA and exhaustive search (ES) is still large. In other words, the approximate function cannot fit the original function well. To reduce the rate performance gap, two high-performance efficient PA methods are proposed to further exploit the rate gain achieved by PA to remove this rate gap. The main contributions of this paper are as below:


	\begin{enumerate}
		
		\item
		
		Given the transmit beamforming vector at BS and the reflection coefficient matrix at IRS are designed well, the SNR expression has been derived to be a function of PA factor $\beta$. To improve the rate performance of conventional gradient ascent (GA), equal-spacing-multiple-point-initialization (ESMPI) is combined with GA to form an enhanced GA, called ESMPI-GA, which is closer to ES than existing methods. Its computational complexity is the order  $\mathcal{O}\{KI\}$ float-pointed operation (FLOPs), where $K$ is the number of equal-spacing initialization points (ESIPs) and $I$ is the average number of iterations per initialization.  According to simulation results, $K$ is larger than or equal to 16, the proposed ESMPI-GA approaches ES in medium-scale or large-scale IRS.


		\item
		Due to the fact that GA has the property of slow linear convergence,  the proposed ESMPI-GA is still high-complexity. To reduce this high complexity, a closed-form  PA method is proposed by using the third-order Taylor expansion (TTE) centered the  pointed $\beta_0=0.5$, called TTE, which is symmetric in the interval $\beta\in(0,1)$ and provide a better approximation than the TPA method in \cite{cheng2023arXiv231009721C}. Afterwards, the Ferrari's method can be utilized to find the roots of the first derivative of  the TTE function of SNR, which forms a candidate set. Simulation results demonstrate that the proposed TTE method harvests about 0.5 bit rate gain over TPA in small-scale IRS with an approximately same order low complexity as TPA  and is closer to the rate performance of ES.
	\end{enumerate}
	
	The remainder of the paper is arranged as follows. In Section II, an active IRS-assisted wireless network system is described and a PA problem is formulated. In Section III, two methods are proposed to solve PA problem. Simulation results are presented in Section IV, and conclusions are drawn in Section V.
	
	$Notations$: Throughout this paper, we denote vectors by lowercase letters and matrices by uppercase letters, respectively. $(\cdot)^T$ and $(\cdot)^H$ stand for transpose and conjugate transpose, respectively.
	Euclidean paradigm, diagonal, expectation and real part operations are denoted as $\|\cdot\|$, diag$(\cdot)$, $E\{\cdot\}$ and $\mathfrak{R}\{\cdot\}$, respectively.

	%

		%
	
	\section{System Model and Problem Formulation}
	\subsection{System Model}
	\begin{figure}[htb]
		\centering
		\includegraphics[width=0.37\textwidth]{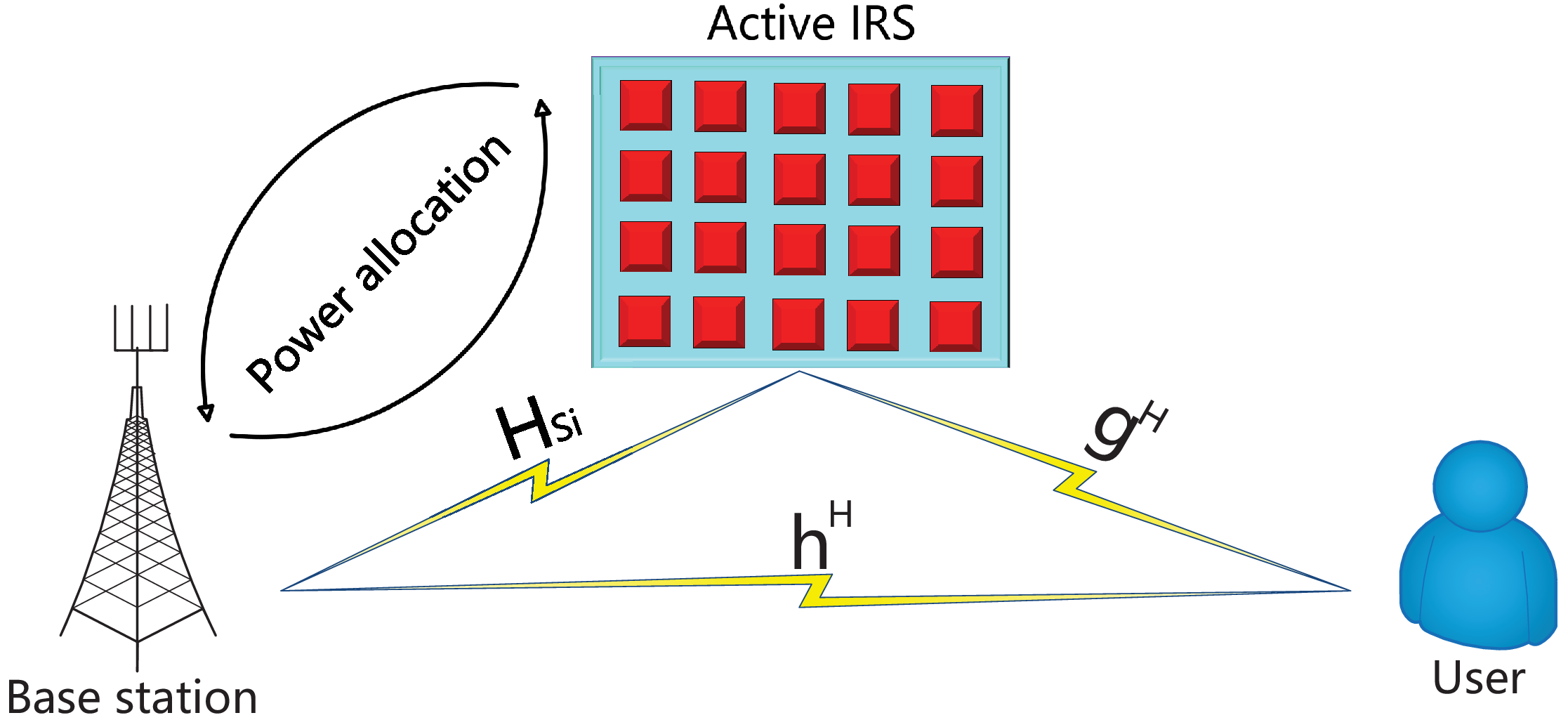}
		\caption{System model of an active IRS-assisted wireless network with PA.}
		\label{systemmodel}
	\end{figure}
	
	Fig.1 sketches an active IRS-assisted wireless communication network with PA, where BS equipped with $M$ antennas serves a single-antenna user with the aid of an  active IRS consisting of $N$ elements. The received signal at IRS is
	\begin{equation}
		\mathbf{y}_{r}=\sqrt{\beta P}\mathbf{H}_{si}{\mathbf{v}}x+\mathbf{n}_I ,\label{y_r1}
	\end{equation}
	where $x$ with $E\{x^Hx\}=1$ is the signal transmitted by BS, $\beta$ is a PA parameter that allocates the power between BS and IRS, $P$ is the total power being the power sum of BS $(\beta P)$ and IRS  $((1-\beta) P)$, ${\mathbf{v}} \in\mathbb{C}^{M\times 1}$ with $\|\mathbf{v}\|_2^2=1$ is the transmit beamforming vector at BS,
	$\mathbf{H}_{si}$ $\in$ $\mathbb{C}^{{N \times M}}$ denotes the channel from BS to active IRS, $\mathbf{n}_{I} \in\mathbb{C}^{N\times 1}$ is the additive white Gaussian noise
	(AWGN) with distribution $\mathbf{n}_{I}\sim\mathcal{C}\mathcal{N}(0,~\sigma^{2}_I\mathbf{I}_N)$.
	The transmit signal at IRS is
	\begin{align}
		\mathbf{y}_t
		&=\mathbf{\Theta}\mathbf{y}_r
		=\sqrt{\beta P}\mathbf{\Theta}\mathbf{H}_{si}{\mathbf{v}}x+\mathbf{\Theta}\mathbf{n}_I,\label{y_t}
	\end{align}
	where matrix $\mathbf{\Theta}=\text{diag}(\alpha_{1} e^{j\theta_1}, \cdots, \alpha_{N} e^{j\theta_N}) \in\mathbb{C}^{N\times N}$ consists of the reflection coefficient of each active IRS element, $\alpha_{n}$ and $\theta_n$ respectively stand for the amplification factor and  phase shift of the $n$-th element. The received signal at user can be represented as
	\begin{align}
		y
		&=\sqrt{\beta P}(\mathbf{g}^H\mathbf{\Theta}\mathbf{H}_{si}+\mathbf{h^H}){\mathbf{v}}x+\mathbf{g}^H\mathbf{\Theta}\mathbf{n}_I+z\nonumber\\
		&=\sqrt{\beta P}(\boldsymbol{\theta}^H\mathbf{G}\mathbf{H}_{si}+\mathbf{h^H}){\mathbf{v}}x+\boldsymbol{\theta}^H\mathbf{G}\mathbf{n}_I+z\nonumber\\
		&=\sqrt{\beta P}(\rho\widetilde{\boldsymbol{\theta}}^H\mathbf{G}\mathbf{H}_{si}+\mathbf{h^H}){\mathbf{v}}x+\rho\widetilde{\boldsymbol{\theta}}^H\mathbf{G}\mathbf{n}_I+z,\label{y1}	
	\end{align}
	where $\mathbf{h}^H \in\mathbb{C}^{1\times M}$ and  $\mathbf{g}^{H} \in\mathbb{C}^{1\times N}$ denote the channels from BS to user and IRS to user, respectively.  $\boldsymbol{\theta}=\rho\widetilde{\boldsymbol{\theta}}=[\alpha_{1} e^{j\theta_1}, \cdots, \alpha_{N} e^{j\theta}]^H$,  $\rho=\|\boldsymbol{\theta}\|_2$,  $\|\widetilde{\boldsymbol{\theta}}\|_2=1$ and $\mathbf{G}=\text{diag}(\mathbf{g}^H)$.
	$z \in\mathbb{C}^{1\times 1}$  is the AWGN with distribution $z\sim\mathcal{C}\mathcal{N}(0,~\sigma^2_n)$. The reflected
	average power at IRS is
	\begin{align}
		P_I&= E\{\mathbf{y}_t^H\mathbf{y}_t\}\label{p_i}=(1-\beta)P\nonumber\\
		&=\beta P\rho^2\|\widetilde{\boldsymbol{\theta}}^H\text{diag} (\mathbf{H}_{si}{\mathbf{v}})\|_2^2+\sigma^2_I\rho^2,
	\end{align}
	which yields
	\begin{equation}
		\rho
		=\sqrt{\frac{(1-\beta)P}{\beta P\|\widetilde{\boldsymbol{\theta}}^H\text{diag} (\mathbf{H}_{si}{\mathbf{v}})\|_2^2+\sigma^2_I}}.\label{p2}
	\end{equation}
	The SNR at the user can be formulated as follows
	\begin{align}
		\text{SNR}
		&=\frac{\beta P\|(\rho\widetilde{\boldsymbol{\theta}}^H\mathbf{G}\mathbf{H}_{si}+\mathbf{h}^H){\mathbf{v}}\|_2^2}{\sigma^2_I\rho^2\|\widetilde{\boldsymbol{\theta}}^H\mathbf{G}\|_2^2+\sigma^2_n}.\label{snr}
	\end{align}
	\subsection{Problem Formulation}
	Under the constraints of total transmit power $P$, the optimization problem of maximizing SNR can be cast as
	\begin{align}\label{SNR-Func1}
		&\max_{\beta,\widetilde{\boldsymbol{\theta}},\mathbf{v}}~~~~~	f(\beta)=\text{SNR}(\beta)
		=\frac{\beta P\|(\rho\widetilde{\boldsymbol{\theta}}^H\mathbf{G}\mathbf{H}_{si}+\mathbf{h}^H){\mathbf{v}}\|_2^2}{\sigma^2_I\rho^2\|\widetilde{\boldsymbol{\theta}}^H\mathbf{G}\|_2^2+\sigma^2_n},\nonumber\\
		&~\text{s.t.}~~~~~~~
		0\le \beta \le 1,~ \|\widetilde{\boldsymbol{\theta}}\|_2^2=1,~ \|\mathbf{v}\|_2^2=1. 		
	\end{align}
	Provided that  $\theta$ and $v$ are designed well, the optimization problem with respect to $\beta$ is cast as follows
	\begin{align}\label{SNR-Func2}
		&\max_{\beta}~~~~~		
		f(\beta)=\frac{u\beta^2+2e\beta\sqrt{l_1\beta^2+l\beta+f}+d\beta }{b\beta+a},\nonumber\\
		&~\text{s.t.}~~~~~~~~0 \le \beta \le 1,
	\end{align}
	where
	\begin{align}
		&\mathbf{M}=\text{diag} (\mathbf{H}_{si}{\mathbf{v}}),~
		a=\sigma_I^2P\|\widetilde{\boldsymbol{\theta}}^H\mathbf{G}\|^2+\sigma_I^2\sigma_n^2,\label{a}\\
		&b=P(\sigma_n^2\|\widetilde{\boldsymbol{\theta}}^H\mathbf{M}\|^2-\sigma_I^2\|\widetilde{\boldsymbol{\theta}}^H\mathbf{G}\|^2),\nonumber\\
		&d=P^2\|\widetilde{\boldsymbol{\theta}}^H\mathbf{G}\mathbf{H}_{si}{\mathbf{v}}\|^2+P\sigma_I^2\|\mathbf{h}^H{\mathbf{v}}\|^2,\nonumber\\
		&e=P\mathfrak{R}\{\widetilde{\boldsymbol{\theta}}^H\mathbf{G}\mathbf{H}_{si}{\mathbf{v}}{\mathbf{v}}^H\mathbf{h}\},~f=P\sigma_I^2,\nonumber\\
		&u=P^2(\|\mathbf{h}^H{\mathbf{v}}\|^2\|\widetilde{\boldsymbol{\theta}}^H\mathbf{M}\|^2-\|\widetilde{\boldsymbol{\theta}}^H\mathbf{G}\mathbf{H}_{si}{\mathbf{v}}\|^2),\nonumber\\
		&l=P^2\|\widetilde{\boldsymbol{\theta}}^H\mathbf{M}\|^2-\sigma_I^2P,~ l_1=-P^2\|\widetilde{\boldsymbol{\theta}}^H\mathbf{M}\|^2.\nonumber
	\end{align}
	Observing the objective function $f(\beta)$  in (\ref{SNR-Func2}) , it is clear that it is a nonlinear function. Maximizing the function is usually addressed by iterative methods like GA. In this paper, we will also design a closed-form solution to the maximization problem of function $f(\beta)$  over interval [0,1].

	\section{Proposed two PA schemes }
	
	To fully explore the rate gain of PA, two high-performance PA schemes, namely ESMPI-GA and TTE, are  developed in this section. The former will enhance the rate performance of conventional GA by introducing ESMPI, while the latter offers an extremely low-complexity high-rate  solution.
	

	%
	%
	%
	%

	\subsection{Proposed ESMPI-GA method}
	First, the original SNR function in (\ref{SNR-Func2})  is uniformly sampled  in the interval [0,1] to  form the following set of initialization points
	\begin{align}
		S_1=\{\beta_{1,0}~\cdots,~\beta_{K,0}\}.
	\end{align}
	With $\beta_{k+1,0}-\beta_{k,0} =1/(K-1)$. Subsequently, the GA is applied to each initialization point.  For the $i$-th iteration of the $k$-th initialization point, the corresponding first derivative  of $f(\beta_{k,i})$ with regard to $\beta_{k,i}$ is defined as follows
	\begin{align}
		f'(\beta_{k,i})
		&=\frac{\partial{f(\beta_{k,i})}}{\partial{\beta_{k,i}}}=\frac{g_1(\beta_{k,i})+g_2(\beta_{k,i})}{(b\beta_{k,i}+a)^2},\label{p26}
	\end{align}
	where
	\begin{equation}
		g_1(\beta_{k,i})=ub\beta_{k,i}^2+\frac{eb\beta_{k,i}^2(2l_1\beta_{k,i}+l)}{\sqrt{l_1\beta_{k,i}^2+l\beta_{k,i}+f}},
	\end{equation}
	\begin{align}
		g_2(\beta_{k,i})=&ad+2au\beta_{k,i}+2ae\sqrt{l_1\beta_{k,i}^2+l\beta_{k,i}+f}\nonumber\\		
		&+\frac{ea\beta_{k,i}(2l_1\beta_{k,i}+l)}{\sqrt{l_1\beta_{k,i}^2+l\beta_{k,i}+f}}.
	\end{align}
	Afterwards, we can obtain $\beta_{k,i+1}$ through GA algorithm as shown below
	\begin{align}
		\beta_{k,i+1}=\beta_{k,i}+pf'(\beta_{k,i}),\label{p27}
	\end{align}
	where $p$ represents the step length. The above iteration process will continue until the termination condition $|\beta_{k,i+1}-\beta_{k,i}|\le \eta$ is satisfied, where $\eta$ is the accuracy. It is assumed that the iteration number is $I_k$. Considering the PA factor is limited to the interval [0,1], we update  the candidate solution by using the following equation
	\begin{equation}
		\widetilde{\beta}_k=
		\begin{cases}
			\beta_{k,I_k} & 0\le\beta_{k,I_k}\le1 ,\\
			0 & \beta_{k,I_k}<1,\beta_{k,I_k}>0.
		\end{cases}
	\end{equation}
	Finally, we  find the optimal solution  by maximizing the original SNR function as follows.
	\begin{equation}
		\beta_{opt}=\argmax_{\widetilde{\beta}_k\in S_{A}}\{f(\widetilde{\beta}_k)\},\label{op1}
	\end{equation}
		where 
	\begin{equation}
		S_{A}=\left\{
		\widetilde{\beta}_1,\cdots,~\widetilde{\beta}_K\right\}.
	\end{equation}

	 The details related to ESMPI-GA method is summarized in algorithm 1, which is displayed as follows. Meanwhile, its computational complexity is $\mathcal{O}\{\sum_ {k=1}^{K}I_k\}=\mathcal{O}\{K\bar{I}\}$ FLOPs, where $\bar{I}=K^{-1}\sum_ {k=1}^{K}I_k$.
	\begin{algorithm}
		\caption{The Proposed ESMPI-GA Method}
		\begin{algorithmic}[1]
			\STATE  Define candidate set $S_1=\{\beta_{1,0},~\cdots,~\beta_{K,0}\}$, and the step length $p$ and the accuracy $\eta$.
			\STATE  Initialize $\beta_{k,i}$, $k=1$, and $i=0$.
			\REPEAT
			\STATE Calculate  the gradient  $f'(\beta_{k,i})$ by (\ref{p26}).	
			\STATE Update $\beta_{k,i+1}$ by (\ref{p27}).
			\STATE Set $i=i+1$.
			\UNTIL $|\beta_{k,i+1}-\beta_{k,i}|\le \eta$ or $\beta_{k,i}\notin(0,1)$.
			\STATE Obtain candidate set $S_{A}=\left\{0,~1,~
			\widetilde{\beta}_1,\cdots,~\widetilde{\beta}_K\right\}$.
			\STATE Calculate the optimal solution of $\beta$ by (\ref{op1}).
		\end{algorithmic}
	\end{algorithm}
	
	\subsection{Proposed TTE solution }
	In the previous subsection,  an enhanced GA is presented and its complexity is proportional to the product of $K$ and $\bar{I}$. Due to a linear convergence speed, the proposed  enhanced GA  is high-complexity. Does there  exist a low-complexity closed-form for $\beta$. Now, let us rewrite the function 	$f(\beta)$ as follows
	
	\begin{align}
		f(\beta)=\frac{u\beta^2+2e\beta+d\beta }{b\beta+a},\label{p28}
	\end{align}
	which is a nonlinear function. Converting the square root in numerator into a low order polynomial will significantly simplify the function $f(\beta)$ to find its maximum value. Let us take our this term and define
	\begin{align}
		q(\beta)=\sqrt{l_1\beta^2+l\beta+f}.\label{p30}
	\end{align}
	The third-order Taylor polynomial in \cite{NA9780538733519} is applied to approximate $q(\beta)$ at point $\beta_0=0.5$ as follows
	\begin{align}
		q(\beta) \approx 	
		&q(\beta_0)+q_1(\beta_0)(\beta-\beta_0)+\frac{1}{2}q_2(\beta_0)(\beta-\beta_0)^2\nonumber\\
		&+\frac{1}{6}q_3(\beta_0)(\beta-\beta_0)^3,\label{p31}
	\end{align}
	with
	\begin{align}
		&q_1(\beta_0)=\frac{2l_1\beta_0+l}{2\sqrt{l_1\beta_0^2+l\beta_0+f}},\label{p91}\\
		&q_2(\beta_0)=\frac{l_1}{\sqrt{l_1\beta_0^2+l\beta_0+f}}-\frac{(2l_1\beta_0+l)^2}{4(l_1\beta_0^2+l\beta_0+f)^\frac{3}{2}},\nonumber\\
		&q_3(\beta_0)=-\frac{3l_1(2l_1\beta_0+l)}{2(l_1\beta_0^2+l\beta_0+f)^\frac{3}{2}}+\frac{3(2l_1\beta_0+l)^3}{8(l_1\beta_0^2+l\beta_0+f)^\frac{5}{2}}.\nonumber
	\end{align}
	Now, substituting the above final approximation (\ref{p31}) into equation (\ref{p28}) yields an approximation to the original objective function as follows
	\begin{align}
		\widetilde{f}(\beta)\approx& \frac{u\beta^2+2e\beta g(\beta)+d\beta}{b\beta+a}
		\approx  \frac{k_1\beta^4+k_2\beta^3+k_3\beta^2+k_4\beta}{b\beta+a},\label{p61}
	\end{align}
	where
	\begin{align}
		&k_1=\frac{1}{3}eq_3(\beta_0),~k_2=eq_2(\beta_0)-e\beta_0q_3(\beta_0),\label{pm2}\\
		&k_3=u+2eq_1(\beta_0)-2e\beta_0q_2(\beta_0)+e\beta_0^3q_3(\beta0),\nonumber\\
		&k_4=2eq(\beta_0)-2eq_1(\beta_0)+e\beta_0^2q_2(\beta_0)-\frac{1}{3}\beta_0^3q_3(\beta_0)+d.\nonumber
	\end{align}
	To obtain the optimal solution, the first derivative of $f(\beta)$ is  set to equal zero
	\begin{equation}
		\widetilde{f}'(\beta)= \frac{\beta^4+A\beta^3+B\beta^2+C\beta+D}{(b\beta+a)^2}=0,\label{pm3}
	\end{equation}
	where
	\begin{align}
		&K_1=3bk_1,~
		K_2=2bk_2+4ak_1,\label{pm4}\\
		&K_3=bk_3+3ak_2,~
		K_4=2ak_3,~
		K_5=ak_4,\nonumber\\
		&A=\frac{K2}{K1},~
		B=\frac{K3}{K1},~
		C=\frac{K4}{K1},~
		D=\frac{K5}{K1}.\nonumber
	\end{align}
	
	In accordance with the Ferrari's method in\cite{Ferrari6457462},  we have the candidate  solutions to (\ref{pm3}) are as follows
	\begin{align}
		&\hat{\beta}_1=-\frac{A}{4}+\frac{\eta}{2}+\frac{\mu_1}{2},~
		\hat{\beta}_2=-\frac{A}{4}+\frac{\eta}{2}-\frac{\mu_1}{2},\label{pm5}\\
		&\hat{\beta}_3=-\frac{A}{4}-\frac{\eta}{2}+\frac{\mu_2}{2},~
		\hat{\beta}_4=-\frac{A}{4}-\frac{\eta}{2}-\frac{\mu_2}{2},\nonumber
	\end{align}
	where
	\begin{align}
		&\alpha_1=\frac{3AC-12D-B^2}{3},~\eta=\sqrt{\frac{A}{4}-B+\gamma},\label{pm6}\\
		&\alpha_2=\frac{-2B^3+9ABC+72BD-27C^2-27A^2D}{27},\nonumber\\
		&\mu_1=\sqrt{\frac{3}{4}A^2-\eta^2-2B+\frac{1}{4\eta}(4AB-8C-A^3)},\nonumber\\
		&\mu_2=\sqrt{\frac{3}{4}A^2-\eta^2-2B-\frac{1}{4\eta}(4AB-8C-A^3)},\nonumber\\
		&\gamma=\frac{B}{3}+\sqrt[3]{-\frac{\alpha_2}{2}+\sqrt{\frac{\alpha_2^2}{4}+\frac{\alpha_1^3}{27}}}+\sqrt[3]{-\frac{\alpha_2}{2}-\sqrt{\frac{\alpha_2^2}{4}+\frac{\alpha_1^3}{27}}}.\nonumber
	\end{align}
	%
	
	Considering $\beta$ is confined to the interval $ [0, 1]$,  the above four candidates have the following new form
	\begin{equation}\label{40}
		\widetilde{\beta}=
		\begin{cases}
			\hat{\beta} & 0\le\hat{\beta}\le1 ,\\
			0 & \hat{\beta}<1,\hat{\beta}>0.
		\end{cases}
	\end{equation}
	In the end,  the optimal solution is given as follows
	\begin{equation}\label{41}
		\beta_{opt}=\argmax_{\beta\in S_{B}}\{f(\beta)\},
	\end{equation}
	where 
	\begin{equation}
		S_{B}=\left\{0,~1,
		~\hat{\beta}_1,~\hat{\beta}_2,~~\hat{\beta}_3,~~\hat{\beta}_4\right\}.
	\end{equation}
		This completes our closed-form derivation for the approximate optimal value of $\beta$. The computational complexity of TTE method is $\mathcal{O}\{6\}$ FLOPs. Obviously, this complexity is far lower than those of ESMPI-GA and ES: $\mathcal{O}\{K\bar{I}\}$ and $\mathcal{O} \{K_{ES}\approx10^4\}$, where $K_{ES}$ is the total search points of ES.  Thus, we have the complexity order:  TTE$\ll$ESMPI-GA$\ll$ES.

			\section{SIMULATION RESULTS AND DISCISSION}
			In this section, we validate the achievable rate performance of the proposed ESMPI-GA method and TTE method. Simulation parameters are  set as follows: the spatial locations of BS, IRS and user are [0 m, 0 m, 0 m], [200 m, 0 m, 35 m], [100 m, 50 m, 0 m], respectively. It is assumed all channels follow Rayleigh fading, and the fading factors of the links from BS to IRS, from IRS to user, from BS to user are defined as 2.3, 2.3 and 2.5, respectively. Additionally, $\sigma_I^2=\sigma_n^2=-100 $ dBm.			
					\begin{figure*}
							\setlength{\abovecaptionskip}{-5pt}
							\setlength{\belowcaptionskip}{-10pt}
							\centering
							\begin{minipage}[t]{0.331\linewidth}
								\centering
					            \includegraphics[width=1.1\textwidth]{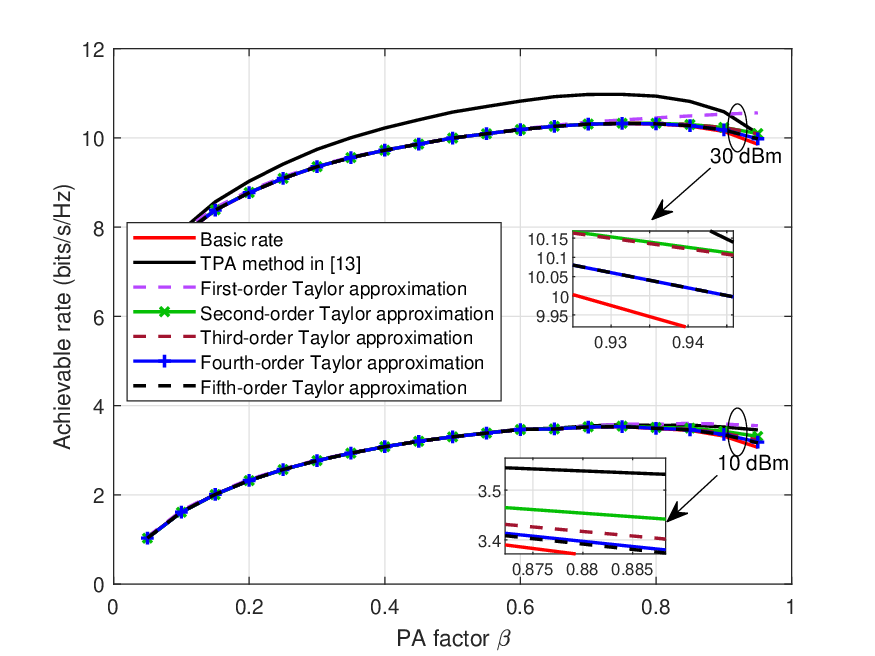}
					            \caption{Achievable rate versus PA factor $\beta$.}
					            \label{Fig-TTE-Appr}
								\end{minipage}
							\begin{minipage}[t]{0.331\linewidth}
									\centering
									\includegraphics[width=1.1\textwidth]{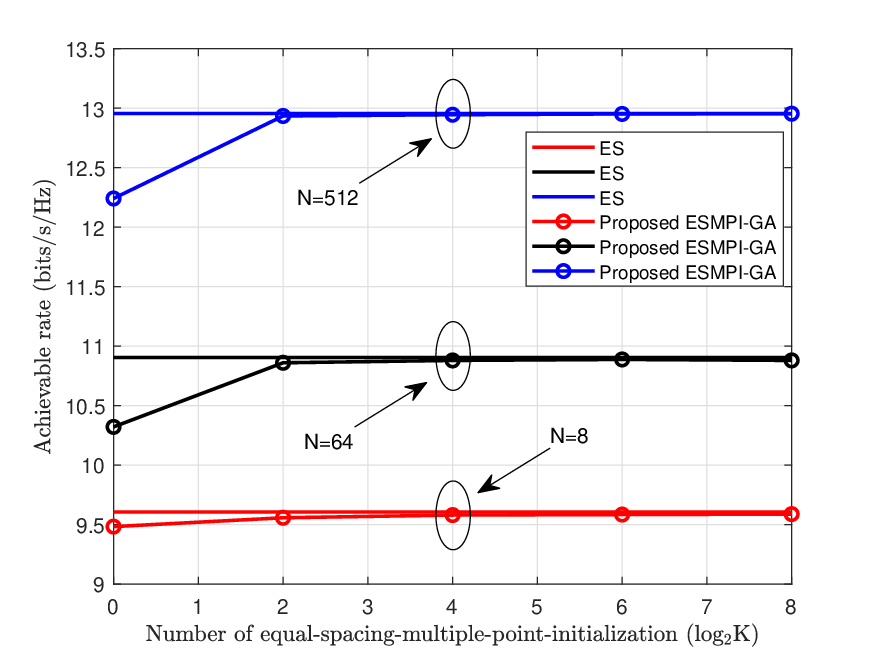}
									\caption{Achievable rate versus the number $K$ of \\ ESI.}
									\label{Conv-Curve}
								\end{minipage}%
							\begin{minipage}[t]{0.331\linewidth}
									\centering
									\includegraphics[width=1.1\textwidth]{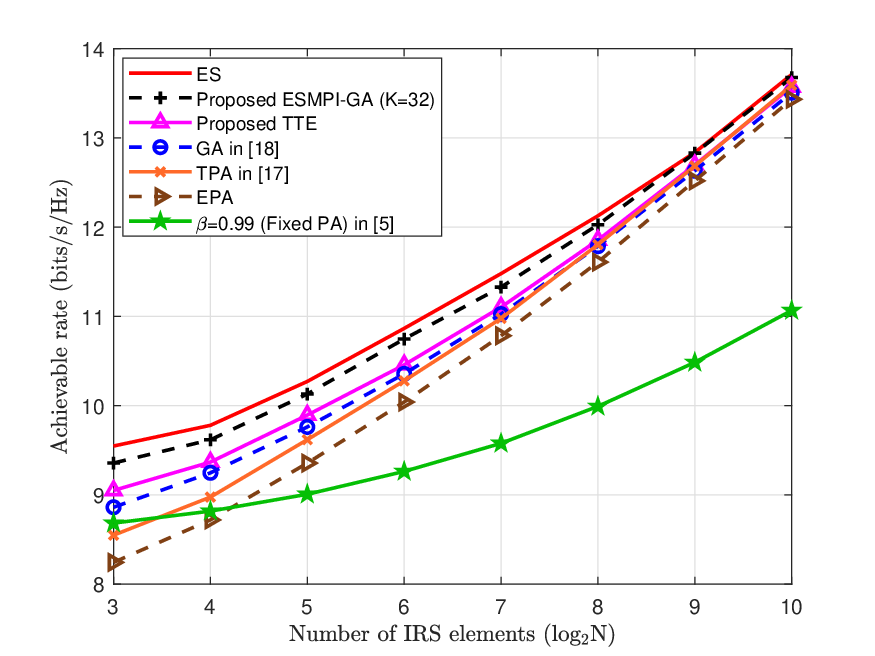}
									\caption{Achievable rate versus the number $N$ of IRS elements. }
									\label{Rate-VS-N}
								\end{minipage}
						\end{figure*}

			Fig. \ref{Fig-TTE-Appr} plots the curves of original rate function and approximate rate
			functions based on distinct order Taylor expansions. From this figure, it is evident that when the polynomial order is larger than 2, the rate gaps between original rate function and approximate rate functions are trivial. Actually, third-order provides a good approximation to the original function curve. Moreover, the proposed TTE is much better than TPA in a large total power constraint.
%
			Fig. \ref{Conv-Curve} demonstrates the convergent curves of  rate versus its number $K$ of initialization points of ESMPI-GA for $N$ = 8, $N$ = 64 and $N$ = 512. 		
			From Fig. \ref{Conv-Curve} , it can be seen that as $K$ increases, the ESMPI-GA method converges to  ES in terms of rate.
			In particular, when $N$  goes to large-scale, the convergent speed of ESI-GA method become faster.  Below,   $K$ is taken to be 32.
			
			Fig. \ref{Rate-VS-N} illustrates the achievable rate versus the number $N$ of IRS elements of the proposed TTE method and ESMPI-GA methods. The porposed two 
			PA methods  ESMPI-GA  and TTE performs much better than  TGA, TPA, fixed PA ($\beta$ $=$ 0.99) in  \cite{DLL9998527} and EPA. 	
			 Additionally, when  the number of IRS elements tends to large-scale, the ESMPI-GA  and  TTE achieve about 2.5-bit rate gain over fixed PA ($\beta$ $=$ 0.99) in  \cite{DLL9998527}. The corresponding rate enhancements are about 22\%. These rate benefits achieved by PA are significant.

			\section{Conclusion}		
			
			In this paper, we have focused on an investigation of PA strategies in an active IRS-aided wireless network.
			First, the expression of SNR has been derived to be a nonlinear function of PA  factor $\beta$, and the corresponding maximization problem  with respect to  $\beta$ has been formulated. Then, two high-performance PA strategies, ESMPI-GA and TTE, have been proposed. In accordance with simulation resulats, the proposed two methods perform much better than existing GA, TPA, EPA  and  fixed PA ($\beta$ $=$ 0.99) in  \cite{DLL9998527}. In paricularly, they may  achieve about $22\%$ rate gain over existing fixed PA method in  \cite{DLL9998527} as the number of IRS elements goes to large-scale.

			\ifCLASSOPTIONcaptionsoff
			\newpage
			\fi

			\bibliographystyle{IEEEtran}
			\bibliography{IEEEfull,reference}
		\end{document}